\input ./style/arxiv-general.cfg
\documentclass[MSNbibl,nameyear,dvips]{arxstspdf}
\makeatletter
   \@ifpackageloaded{graphicx}{}{\usepackage{graphicx}}
\makeatother
\usepackage{flushend}
\usepackage{stfloats}

\volume{30}
\issue{2}
\pubyear{2015}
\firstpage{258}
\lastpage{267}
\doi{10.1214/14-STS497} 
\docsubty{FLA}


\begin{document}
\begin{frontmatter}

\title{A Conversation with Robert C. Elston}
\runtitle{A Conversation with Robert C. Elston}

\begin{aug}
\author[a]{\fnms{Gang}~\snm{Zheng}},
\author[b]{\fnms{Zhaohai}~\snm{Li}\ead[label=e2]{zli@gwu.edu}}
\and
\author[c]{\fnms{Nancy L.}~\snm{Geller}\corref{}\ead[label=e3]{gellern@nhlbi.nih.gov}}
\runauthor{G. Zheng, Z. Li and N.~L. Geller}

\affiliation{National Heart, Lung and Blood Institute, George Washington University and National Heart, Lung and
Blood Institute}

\address[a]{Gang Zheng was a Mathematical Statistician, Office of Biostatistics
Research, National Heart, Lung and Blood Institute, Bethesda, Maryland
20892-7913, USA. He passed away on January 9, 2014, without completing
this.}
\address[b]{Zhaohai Li is a Professor of Statistics and Biostatistics and the
Chair, Department of Statistics, George Washington University, Rome Hall,
5th Floor, 801 22nd St. NW, Washington,
DC 20052, USA \printead{e2}.}
\address[c]{Nancy L. Geller is the
Director of Office of Biostatistics Research, National Heart, Lung and
Blood Institute, 6701 Rockledge Drive, Bethesda, Maryland 20892-7913, USA
\printead{e3}.}
\end{aug}

\begin{abstract}
Robert C. Elston was born on February 4, 1932, in
London, England. He went to Cambridge University to study natural science
from 1952--1956 and obtained B.A., M.A. and Diploma in Agriculture (Dip Ag).
He~came to the US at age 24 to study animal breeding at Cornell University
and received his Ph.D. in 1959. From 1959--1960, he was a post-doctoral
fellow in biostatistics at University of North Carolina (UNC), Chapel Hill,
where he studied mathematical statistics. He~then rose through the academic
ranks in the department of biostatistics at UNC, becoming a full professor
in~1969. From 1979--1995, he was a professor and head of the Department of
Biometry and Genetics at Louisiana State University Medical Center in New
Orleans. In 1995, he moved to Case Western Reserve University where he is a
professor of epidemiology and biostatistics and served as chairman from
2008 to~2014. Between 1966 and 2013, he directed 42 Ph.D. students and
mentored over 40 post-doctoral fellows. If one regards him as a founder of
a pedigree in research in genetic epidemiology, it was estimated in 2007
that there were more than 500 progeny. Among his many honors are a NIH
Research Career Development Award (1966--1976), the Leadership Award from
International Society of Human Genetics (1995), William Allan Award from
American Society of Human Genetics (1996), NIH MERIT Award (1998) and the
Marvin Zelen Leadership Award, Harvard University (2004). He is a Fellow of
the American Statistical Association and the Institute of Mathematical
Statistics as well as a Fellow of the Ohio Academy of Science. A~leader in
research in genetic epidemiology for over 40 years, he has published over
600 research articles in biostatistics, genetic epidemiology and
applications. He has also coauthored and edited 9 books in biostatistics,
population genetics and methods for the analysis of genetic data.

The original conversation took place on August 4, 2009, during the Joint
Statistical Meetings (JSM) in Washington, DC by GZ and ZL. NLG had dinner
with RCE during the 2013 JSM in Montreal, Canada, and added supplementary
material and edited the conversation. RCE updated and clarified certain
points.\vspace*{1pt}
\end{abstract}

\begin{keyword}
\kwd{British statisticians}
\kwd{family studies}
\kwd{genetic epidemiology}
\kwd{history of statistical genetics}
\kwd{pedigree data}
\kwd{S.A.G.E. software}
\kwd{statistical genetics}
\kwd{statistics biographies}\vspace*{1pt}
\end{keyword}
\end{frontmatter}

\section{Early Education}

\textbf{Gang and Zhaohai:} Robert, it is a great pleasure to have this
opportunity to talk with you about your life, research, career, mentorship
and some of your views of genetic epidemiology.

Can you begin by telling us about your early years?

\textbf{Robert:} I was born in London, and I was 7 years old when World War
II broke out (1939). My brothers and I were evacuated to a little village,
Lea Green in Hertfordshire, about 30 miles from London. That's where I
first loved farming and thought I'd be a farmer when I grew up. In 1941, my
father arranged for us to live in Hertford, where Battersea Grammar school
had been evacuated from London. I don't know how he got me into that
grammar school since I was really too young. They put me in the lowest form
(grade). I eventually took what was called the school certificate at 14
while most took it when they were at 16.

\textbf{Zhaohai:} What did you study in high school?

\textbf{Robert:} In addition to the usual subjects, we studied French and a
year later got to choose Latin or German. I did Latin mainly because my
elder brothers had done Latin, and because I knew I needed Latin to go to
Oxford or Cambridge. The following year, the class master who taught Latin
chose the two or three best students and said: Okay, you will do Greek. My
brothers had done Greek but they had to give up physics to do Greek and I
was not going to give up physics! I said, if you want me to do Greek, I'll
need to eliminate history or geography or both. So they agreed!

When World War II ended, Battersea Grammar school moved back to London, so
I had one year at Hertford Grammar school. In 1946, we returned to London
as a family. Then I went to University College School, which was ``a public
school,'' meaning it was open to anyone who was willing to pay (laughs).
Although I studied Latin and Greek and the classics for two years, I also
wanted to do science. As I was young, I could stay there in the sixth form
for four years, mixing 2 years of classics (Latin, Greek, a little French
and ancient history) with 2 years of biology, physics and chemistry. I
never studied calculus, highly unusual in the US, but not so unusual at
that time for a science student in England. When I got to the states, I
estimated that what we did in England in the sixth form was equivalent to
one or two years of undergraduate work in the states.

\section{Cambridge University}

\textbf{Gang:} How did you get to Cambridge University?

\textbf{Robert:} In those days, the way you got to a university was\vadjust{\goodbreak} either
you were rich or royalty, or you sat for a scholarship examination. I
applied to both Reading and Cambridge Universities and had to take a
scholarship examination at each. The scholarship wasn't much in terms of
the money, but if you passed the scholarship exam, the local government
would pay for your education.

\textbf{Zhaohai:} So you got the scholarship?

\textbf{Robert:} No, I actually failed. For Reading University, the three
subjects I chose for the exam were Greek, French and chemistry. They
thought that combination quite useless. At Cambridge, there were six of us
competing for one scholarship. At that time, the School of Agriculture at
Cambridge had a three-year Bachelor's degree in agriculture. I had already
decided to do something in agriculture. But I had taken a special
scholarship exam at Magdalene College for people who would spend four
years. The first two years would be Part I of the natural science tripos.
[A tripos is the course system at the University of Cambridge.] Then at the
end of the second year, I would have a choice, either a two year diploma in
agriculture or continue with Part II of the natural science tripos followed
by a one year diploma in agricultural science. Although I didn't get that
scholarship, I did well in the exam, and they said they would accept me
into Magdalene College for the four-year program in two years' time, with
the government giving me some support. Why two years? Because they ``knew''
I would have to serve two years in the military. So I had those two years
to spend. Since I was going to study agriculture, I was able to get a
deferment from the military to work on a farm, but I remained eligible for
military service and could be called up later. So I worked on a farm for a
year and then spent my next year in France where I perfected my French (my
mother's mother tongue).

\textbf{Gang:} What happened after those two years?

\textbf{Robert:} I returned to England and went to Cambridge. For the
natural science tripos, I spent two years doing work for Part I. I had to
have three science subjects. My original idea was to do botany, zoology and
chemistry (organic and biochemistry). After one year, I really didn't like
botany and decided I wanted to do mathematics. So I changed from full
subject botany to half subject botany and half subject mathematics. I had
to teach myself calculus, which I did with a little book called \textit{Calculus
Made Easy} by Silvanus P. Thompson (\citeyear{Tho46}). Clearly, I needed private
tutoring for mathematics. Wally Smith (Walter Laws Smith), whom I knew from
my extracurricular activity on the stage (we were both members of the
Pentacle club, which was a magic club), became my mathematics supervisor.
He told me I wasn't very good at mathematics, but I stuck it out! I know
mathematics lowered my exam result at Cambridge!

\textbf{Zhaohai:} What did you do next? Did you get to do genetics at
Cambridge?

\textbf{Robert:} My first choice for a Part 2 tripos was biochemistry,
which I really enjoyed, but between the lectures and the labs, the hours
were too long, so I went to my tutor for advice. I said I was thinking
about Part 2 genetics. I remembered his words so well, ``You know this
program in genetics here is new. And this man [R.~A.] Fisher is considered
eccentric by some, and it may not stand you in good stead in later life for
it to be known you worked with him.'' That is why I did not do a year with
Fisher! So I ended up doing the two-year diploma in agriculture.

\textbf{Gang:} Who taught you statistics at Cambridge?

\textbf{Robert:} I had lectures from four people. Dennis Lindley gave us
three weeks on statistics as part of the half-subject mathematics. He
taught me significance testing. He did not believe in it but he taught it!
In the same year, I did have lectures from Fisher because those who did
zoology could do an optional series of lectures with Fisher on genetics. So
I did do genetics with Fisher. I'll tell you a joke he told (translated
into American). I did not know what he was doing with adding, subtracting
and dividing for a 2 by 2 table. He came up with this number. He said,
``Now, I am going to call this number chi-squared. Don't be alarmed. I know
you are all biologists. It is no worse than calling a dog `Lassie'!'' He
said, ``If this number is greater than~4, perhaps there is something going
on.'' This was in a lecture hall which could hold 200 people. At the first
lecture, there might have been 150 people; second lecture, 50; third, about
15! He made us all sit in the front row. He could sense his audience. I
don't remember exactly what he said, but it was something like: if I say
something is always transmitted from mother to daughter, then clearly it is
never transmitted from mother to son. Then he said, ``I do hope I'm not
making a mistake in logic. Do stop me if I make a mistake in logic, won't
you?'' He could sense that we were all thinking about the truth of what he
had said!

While I was spending a year in France, I had read Fisher's book for
research workers (\citeyear{Fis50}) and his book on experimental design (\citeyear{Fis51}). So one
day after lecture I said to Fisher, I read your book. What is the
difference between a standard deviation and a standard error? He looked me
up and down and said: ``Your height is a deviation from the mean. It is not
an error.''

Then I also had lectures with Anscombe and R.~C. Campbell, because they were
teaching agriculture students. There we learned about experimental and
split-plot design, and basic statistics. We also knew how to calculate
F-statistics using a hand calculator. And I had lectures from Wishart.
Wishart used a little book he had written with Sanders called \textit{Principles
and Practice of Field Experimentation} (\citeyear{WisSan55}). He taught us how to lay out
plots in the field and for agricultural experimental designs. We were just
agriculture people! He was in the School of Agriculture.

\textbf{Zhaohai:} And you still wanted to be a farmer after you got your
diploma in agriculture?

\textbf{Robert:} What else was I going to do? I learned how to run a farm
in England and knew quite a lot of animal physiology, plant physiology,
soil science, how to work out the feeding and animal nutrition, but I had
no capital. To be a farmer, you needed capital for the land and the
machinery. I couldn't afford a farm.

\section{Coming to the  US  for a Ph.D.}

\textbf{Zhaohai:} How did you end up coming to US for your Ph.D. in animal
breeding?

\textbf{Robert:} I got the B.A. in 1955. The way it worked was that two
years after you got your B.A., you could pay 10 guineas and you got an M.A.
So I had an M.A. My mother wanted me to take an academic job. I saw a
notice: Fellowships to America. All right, I thought, I'll just go to
America for just one year. These were King George VI memorial fellowships
from the English Speaking Union of the US and they were giving about 25
scholarships a year. You could have up to three choices of where you wanted
to go, but you had to sign that you would go wherever they sent you. My
choices were UC Davis where Michael Pease was doing chicken breeding or
Ames, Iowa, which was known for dairy cow breeding. I left the third choice
blank. They sent me to Cornell, where there was a department of animal
husbandry. There I was sent to Chuck (Charles Roy) Henderson, who said all
his students minored in biometry, and suggested that I go to see Professor
Federer. So I went to see Walter Federer, who asked me why I did not stay
for a Ph.D. I said I had only money for a year. He told me that they would
find me money.

I had to return to England at the end of the year because I was called for
military service; and if I passed my 26th birthday outside of
England, I could have been called up to age 36. Again I avoided military
service, this time by working on a pig farm. I was able to leave England
before my 26th birthday because the farm owner was willing to say I
was still there. I actually spent my 26th birthday on the high seas
en route back to Cornell. I did this so I could get to Cornell when the
semester began.

So I returned to study animal breeding for a Ph.D. with Chuck Henderson
with minors in biometry and mathematics. My thesis was in mixed model
nonorthogonal ANOVA. We had one of the first computers, an IBM 650. I spent
three months with punch cards to invert a $79\times 79$ matrix!

\textbf{Zhaohai:} Who were your contemporaries in graduate school at that
time?

\textbf{Robert:} I was exactly contemporaneous with a student of Chuck
Henderson's, Shayle Searle. He had a degree in mathematics and a diploma in
statistics. Chuck Henderson had just spent a year at the New Zealand Dairy
Board with Shayle Searle and recruited him to be a graduate student at
Cornell. I learned a lot of from both of them. Chuck Henderson told me he
was an animal scientist, not a statistician. The reason he was doing BLUP
(Best Linear Unbiased Prediction) was because statisticians wouldn't do it
for him. He~never considered himself as a statistician at all.

\section{From Cornell to University
of North Carolina at Chapel
Hill (UNC)}

\textbf{Gang:} How did you choose a post-doctoral fellowship in statistics
after finishing up your Ph.D. in animal breeding?

\textbf{Robert:} I was going to finish my Ph.D. in the summer of 1959. I
didn't know what to do next. Walter Federer advised me to do a
post-doctoral fellowship in statistics. He said I should apply to
Princeton, where there were fellowships in statistics for biologists. They
paid \$5000 a year, tax free. So I applied. Before I heard from Princeton,
I drove to Miami for an international student conference. On my way back, I
stopped at Chapel Hill to see my old friend Wally Smith, who had moved to
the Department of Statistics there. He told me they could offer me \$4,800
a year as a tax-free fellowship in the Department of Biostatistics, as
Bernard Greenberg, the chairman, had some money. So I went to UNC at Chapel
Hill. I wrote to Princeton that I was no longer interested, and received
the nicest letter back from Sam Wilks. From that experience, I learned that
when you are trying to recruit students or post-docs, write a nice
letter.

\textbf{Zhaohai:} How long was your post-doc at Chapel Hill?

\textbf{Robert:} Just one year. Bernie Greenberg insisted that teaching was
part of the training for all students, pre- and post-docs, so I became a
teaching assistant for Statistics 101 for public health. My affiliation was
with biostatistics, but I had my office in the statistics department. In
addition, I took 5 theoretical statistics courses in each of the two
semesters, although I did not do all the homework. I took courses in
multivariate analysis from S.~N. Roy and Norman Johnson, response surface
designs from R.~C. Bose, experimental design from Indra Chakravarti and
David Duncan, nonparametric statistics from Wassily Hoeffding
(U-statistics) and mathematics for statistics from Wally Smith. During that
time, James Durbin, Maurice Kendall and E.~J. Hannan (time series) were
visitors. I also published work from my dissertation, my first paper in
\textit{Biometrics} (\citeyear{Els61}).

\begin{figure*}

\includegraphics{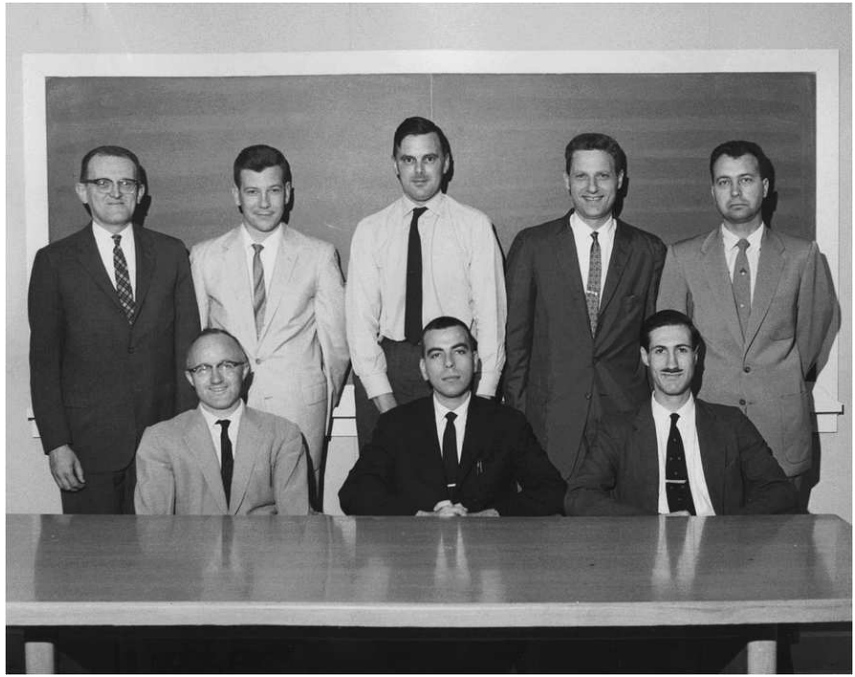}

  \caption{The department of biostatistics at UNC, Chapel
Hill circa 1960. Front row (left to right): Thomas Donnelly,
Bernard Pasternack, Robert Elston. Back row (left to right): Roy
Kuebler, Jr., James Grizzle, David Newell (visiting),
Bernard Greenberg, Bradley Wells.}\label{f1}
\end{figure*}

\textbf{Gang:} You spent most of 1959 to 1979 at UNC Chapel Hill. Since you
came as a post-doctoral fellow, how did you manage to stay?

\textbf{Robert:} In order for me to stay at Chapel Hill, Bernie Greenberg
suggested a job as a Research Assistant Professor of Pathology to work on a
project for the blood bank, which he thought must be related to genetics,
my interest. I worked on a project to estimate the amount of blood that the
blood bank should keep on hand (Elston, \citeyear{Els62}; Elston and Pickrel,
\citeyear{ElsPic63}). I
simulated blood units being purchased by the blood bank and being sent out
for use (Elston, \citeyear{Els62}).

Computers and statistics had not been used before in blood banking.
Analyzing six months of the blood bank's records, I found I could fit a
negative binomial distribution to the number of units that came into the
bank for seven of the eight major blood types, but not for the O negative
blood data. That was because O negative blood can be transfused into anyone
and so O negative donors were often requested to donate blood, rather than
donating simply at random. The blood bank director was impressed with that
finding purely by statistical analysis. I ended up writing several other
papers on the blood bank project (Elston and Pickrel, \citeyear{ElsPic65};
Elston, \citeyear{Els66,Els68,Els70}).

Toward the end of that second year, Bernie Greenberg said he needed someone
to teach bioassay the next year, and did I know anything about it? I said I
knew a little, and yes, I could teach that, but in truth, I was one chapter
ahead of the students most of the time.

\textbf{Gang:} But then there was this two year gap, 1962--64 when you went
to Aberdeen. Why was that?

\textbf{Robert:} My third year at Chapel Hill was the sixth year I had been
in the US. I had a J-1 visa and US law required that you had to return to
your home country for at least two years. Exceptions to stay in the US were
only by an act of Congress. David Finney contacted me for a permanent
position as a Senior Biometric Fellow in Aberdeen Scotland and, since I
couldn't stay in the US, I accepted. My wife and I didn't like the idea of
going to Aberdeen very much, but this was a permanent position. (I had just
got married in Chapel Hill and my wife came from Gloucestershire, 100 miles
west of London, so we were both from southern England.) After being in
Aberdeen for about six months, I put down a deposit on a house, and the
next day I got a letter from Bernie Greenberg asking me to come back to
Biostatistics as an Associate Professor after the required two years
outside the US. He asked me what salary I would want and when I named the
largest salary I dared, he offered me 25\% more. I was trapped! This time I
came to the US with a green card. We returned to Chapel Hill with a
nine-month old daughter, and for her to get her visa, I had to sign on her
behalf that she wasn't coming into the US for the purpose of becoming a
prostitute; she remarked recently that she kept her half of the bargain!

\textbf{Gang:} Describe the Biostatistics Department on your return.

\textbf{Robert:} I was the 6th or 7th faculty member of the
department. The department grew with the help of federal grant
support. In the mid-sixties, we were tremendously successful. In 1966, I
managed to get a five-year Career Development Award and then a five-year
renewal. There was an interdepartmental training grant in genetics and
biostatistics had its own training grant, but there was no Ph.D. in
biostatistics. At that time, the Ph.D. students funded by the departmental
training grant took either a Ph.D. in Experimental Statistics at Raleigh or
a Ph.D. in Statistics at Chapel Hill, with a minor in Public Health. I
wanted a Ph.D. program in biostatistics with a minor in genetics to have
students funded by the interdepartmental training grant, so I wrote the
Ph.D. proposal. Greenberg was told it wasn't broad enough, so I rewrote it
allowing for minors in genetics, demography and other fields as well. The
Ph.D. program in Biostatistics officially began in 1968 with Rose
Gaines-Das being the first to get a Ph.D. in biostatistics, with a thesis
in statistical genetics.

Initially, it was difficult for my students to get positions. That's why
Joe Haseman went to the National Institute of Environmental Health
Sciences; there were no academic positions in statistical genetics. Haseman
could not get a job in statistical genetics despite the fact that Haseman
and Elston (\textit{Behavior Genetics}, \citeyear{HasEls72}, from Haseman's dissertation)
became the most cited paper ever published in \textit{Behavior Genetics}.

\begin{figure*}[b]

\includegraphics{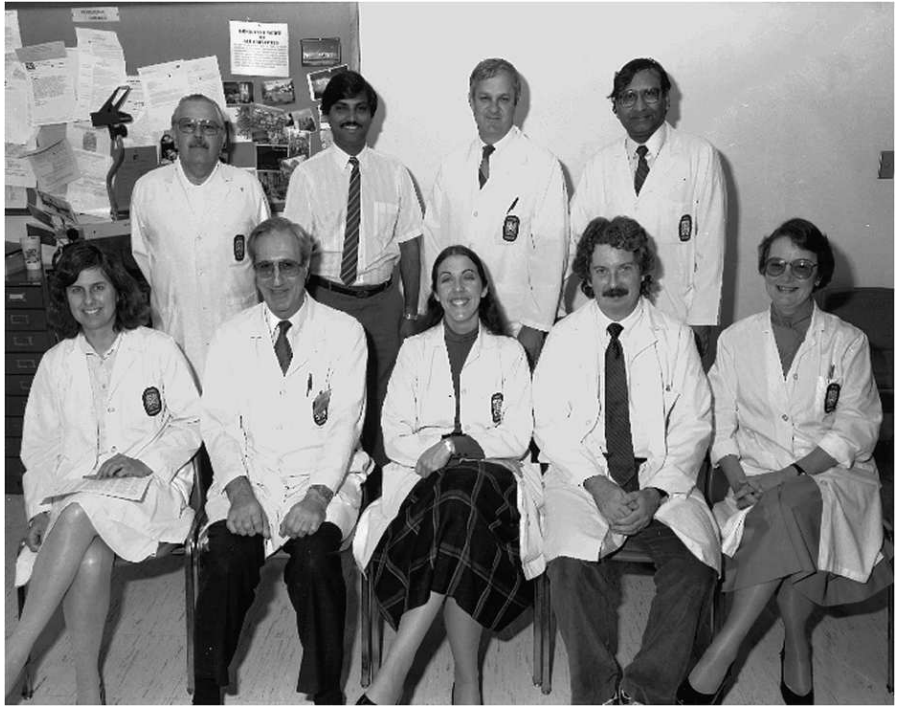}

  \caption{Faculty of the Dept. of Biometry and Genetics,
LSUMC, 1990. Front row (left to right): Bronya Keats, Robert
Elston, Joan Bailey-Wilson, Alexander Wilson, Mary Kay Pelias. Back row
(left to right): Miguel (Mike) Guzman, Varghese
George, William Johnson, Yogesh Patel.}\label{f2}
\end{figure*}

\begin{figure*}[b]

\includegraphics{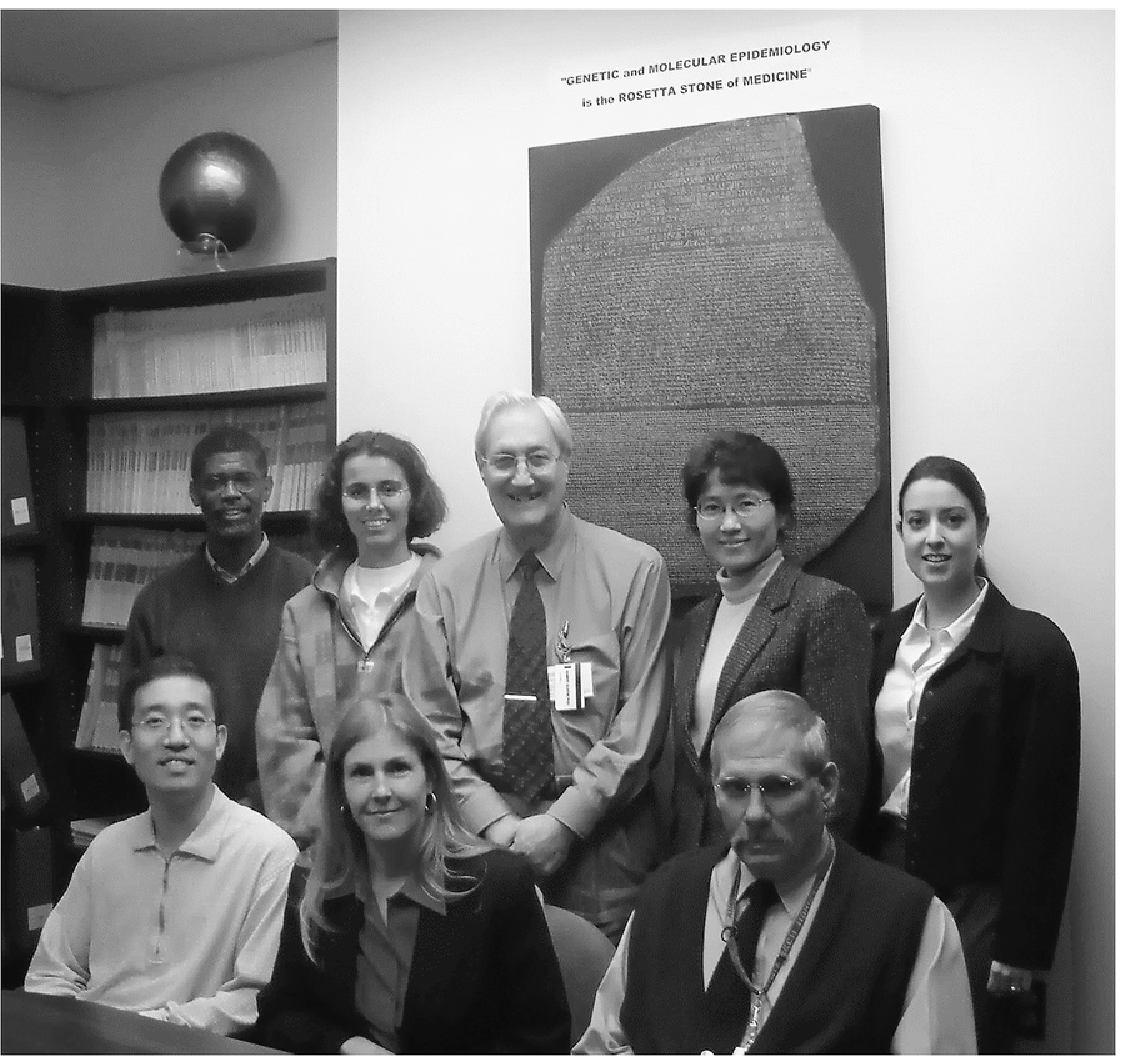}

  \caption{Professor Elston with students in the Division of
Molecular and Human Genetics, Department of Epidemiology and Biostatistics,
CRWU, 2004, with British Museum replica of the Rosetta Stone in the
background. The caption above reads, ``Genetic and molecular epidemiology
is the Rosetta stone of medicine.'' Front row (left to right):
Kijoung Song, Denise Daley, James Schick. Back row (left to right): Ronald Blanton, Murielle Bochud, Robert Elston, Danghong
Song, Courtney Gray-McGuire.}\label{f3}
\end{figure*}

By now, many of my Ph.D. students funded on that training genetics grant
are retired. I can't imagine why!

\textbf{Gang:} While you were at UNC, you did a lot of traveling. How did
you manage that?

\textbf{Robert:} With the Career Development Award, my position didn't cost
the university and Bernie Greenberg said I could do whatever I wanted
because it didn't cost him anything. This allowed me to visit the
University of Hawaii to work with Newton Morton and D.~C. Rao for one year
and, during summers I had further trips to Hawaii and England (the Galton
Laboratory in London and the University of Cambridge). When I visited the
Galton laboratory (1967), I met John Stewart, who was a graduate student at
Cambridge. We ended up writing a paper together, which appeared in
\textit{Human Heredity} (\citeyear{ElsSte71}), a minor journal at that time. We computed
the likelihood of the model for the observed phenotype data in a given
pedigree. We could handle large pedigrees and relatively few markers. I~didn't then know I was using Bayes' theorem recursively to compute the
likelihood. Stewart's contribution was to apply the result to linkage. In
the discussion, Stewart wrote that this paper answered a fundamental
question in human genetics, that is, is some phenotype polygenic or is
there a major gene? It was Ken Lange who named this ``the Elston--Stewart
algorithm.'' It was overall a most productive time.

\section{From UNC to Louisiana State
University Medical Center (LSUMC)}

\textbf{Gang:} Why did you leave UNC for LSU in 1979?

\textbf{Robert:} I moved to LSU for two major reasons. I went to New
Orleans for the ENAR meeting and they wanted me to come there to be chair
of the Department of Biometry in the LSU Medical Center. They offered me a
hard money position. That was the first reason: all positions at UNC were
soft money and by then I had four children, all within six years of age, to
put through college. The second reason was that at UNC I had gotten a grant
which allowed me to purchase my own computer and the university would only
let me house it in the computer center. At that time, nobody in the School
of Public Health was permitted to have his own computer. So those were the
primary reasons that I left.

\textbf{Zhaohai:} Tell us about your years at LSU.

\textbf{Robert:} Even though I had a hard money position, I kept writing
grants. Because I had all of these federal grants, I was able to start a
Ph.D. program in statistical genetics and expand the faculty. I wrote four
proposals for Ph.D.s and masters' degrees in Biometry and Genetics. Alec
Wilson, Joan Bailey-Wilson and George Bonney became part of my faculty.

\textbf{Gang:} What kind of training did you give at LSU?

\textbf{Robert:} I trained several post-docs there. I especially like to
train statisticians to do genetics. At LSU, I had a training grant from
NHLBI which was initially only to train post-docs. From 1992--1993, Dan
Schaid of the Mayo Clinic was my post-doc. I remember the year because
Hurricane Andrew hit Louisiana that year. It was supposed to hit New
Orleans. We boarded up the windows and left for our eldest daughter's
wedding in Ann Arbor. We thought we might not have a house when we got
back. But the hurricane missed New Orleans. Dan Schaid completed his year
with me and went back to the Mayo Clinic and was able to analyze the
genetic data that they had been collecting. He is now a leader in the field
of genetic epidemiology.

\textbf{Zhaohai:} Why did you leave LSU?

\textbf{Robert}: My faculty was good and I wanted to raise their salaries.
The administration said there were no faculty raises and ``no exceptions.''
Of course, there were exceptions! That's why I'm a lousy administrator: I
refuse to lie! In one of the following years, the Chancellor wrote a letter
to the department heads saying that again there would be no faculty raises
and noted that good people would leave and ``this should be taken as an
opportunity.'' I was fed up with being a department head anyway and had
\$1,000,000 in grant money. That and the climate were the reasons I decided
to leave!

\section{From LSU to Case Western
Reserve University (CWRU)}

\textbf{Zhaohai}: You have been at CWRU since 1995. Why did you choose
CWRU?

\textbf{Robert:} My wife hates the heat, so staying in the south was out of
the question. She wanted to go to Maine or Vermont, so Cleveland was a
compromise. I accepted a full professorship at CWRU without administrative
responsibilities so I could get some work done!

\textbf{Gang:} Tell us about the department when you arrived.

\textbf{Robert:} I was hired by an epidemiologist, Alfred Rimm. He wanted
me to have my own division, so we called it Genetic and Molecular
Epidemiology.  The department is really a mini-school
of Public Health. Aside from my division, it had divisions of Epidemiology,
Biostatistics, Health Services Research and Public Health. The names have
changed over the years, but with the exception that there are no longer
formal divisions, the structure is the same.

When I moved, only two people from LSU came with me, Xiuqing Guo, a
graduate student, and Hemant Tiwari, a post-doc. I was also able to take my
training grant in biometric genetic analysis because nobody remained at LSU
who could do the work. Joan Bailey-Wilson and Alec Wilson moved to the
National Human Genome Research Institute (NHGRI) of NIH because they had
family nearby, in Baltimore. I was also able to take my computers.

Al Rimm asked me to do genetics only, not biostatistics and I did that for
over ten years. One project was S.A.G.E. (Statistical Analysis for\break Genetic Epidemiology), which I had started in New Orleans, funded by an NIH
Resource Grant. I also took that with me. The Resource Grant required
collaborations, providing a service for which you had to charge (the
S.A.G.E. software), training and dissemination (S.A.G.E. courses).
Initially, there was a charge for S.A.G.E. because the grant required that.
Beginning in 2005, we were able to distribute S.A.G.E. for free (see Elston
and Gray-McGuire, \citeyear{ElsGra04} and \url{https://code.google.com/p/opensage/}).
 Version
6.2 was meant to be web based so people could use other programs with it,
but funding to complete this project never materialized.

\textbf{Gang:} How come you became department chair at CWRU?

\textbf{Robert:} In 2008, Al Rimm resigned as chair and they asked me to be
interim department chair and I agreed. They needed a real chair to apply
for stimulus money, so in 2009, they took the ``interim'' away. They
continued to advertise for a real chair and it took several years---until
now (2014)---to fill the position. Call it a second childhood!

\begin{figure}

\includegraphics{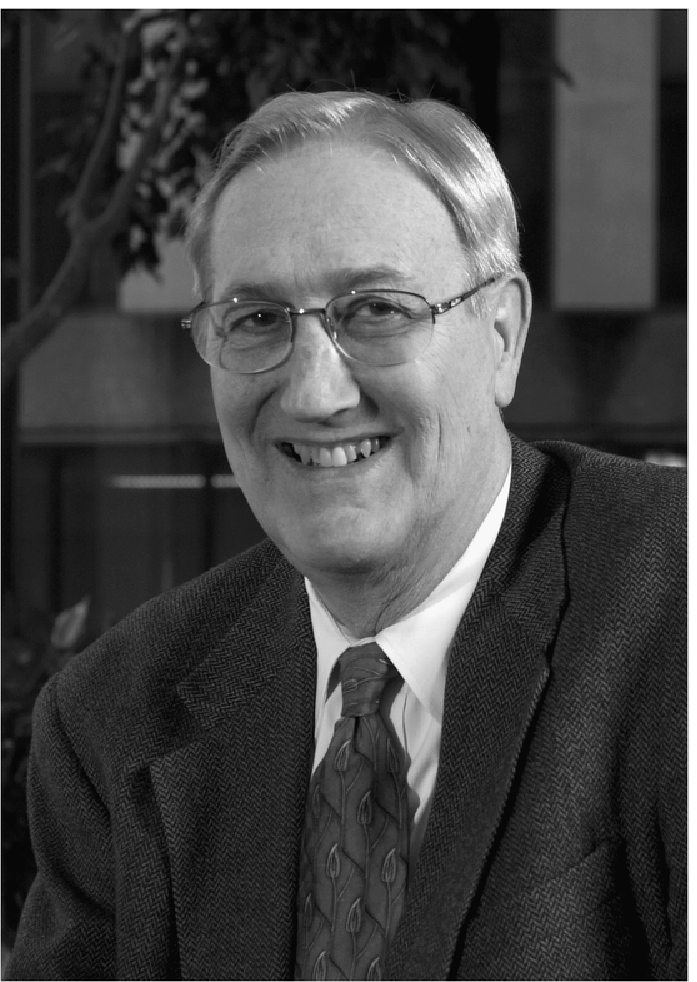}

  \caption{Robert Elston at Case Western Reserve University,
2007.}\label{f4}\vspace*{6pt}
\end{figure}

\textbf{Zhaohai:} How did you arrange your time as chair on administration,
research and mentoring graduate students and post-docs?

\textbf{Robert:} When I first became chair, we needed to reorganize our
Ph.D. program. Under the previous chair, the department was acting as
though each division had a separate Ph.D. program. The different divisions
found it hard to agree on one Ph.D., but the graduate school did not
recognize multiple Ph.D. programs in one department. So I put a lot of
effort into establishing the one Ph.D. program in Epidemiology and
Biostatistics, with several different concentrations. When that was done,
with several faculty committees to make sure that the program ran smoothly,
a lot of the administration was taken care of. And I have always considered
my research and mentoring of students, both pre- and post-docs, to be all
part of one and the same thing.

\begin{figure*}

\includegraphics{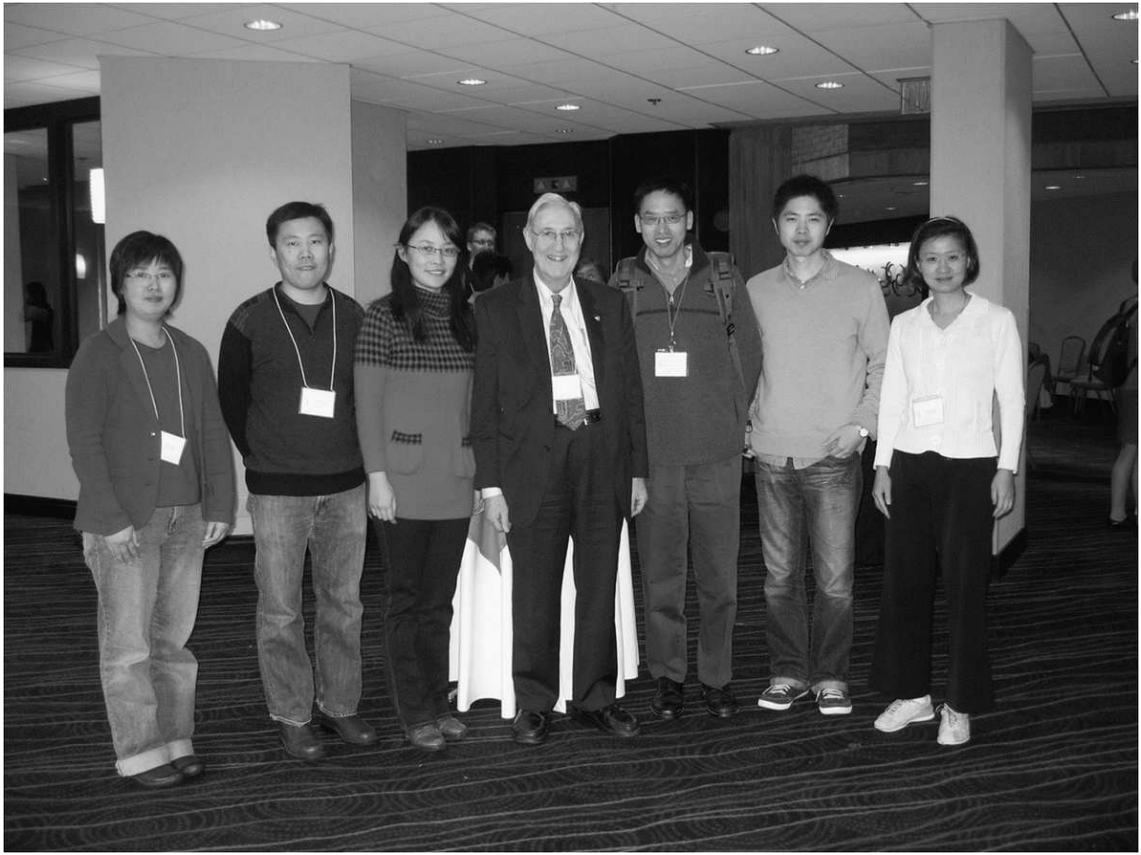}

  \caption{Students with Professor Elston at a meeting of the
International Genetic Epidemiology Society in Boston, August, 2007. Left to
right: Wei Guo, Qing (Jerry) Lu, Lu Zhang, Robert Elston,
Xiaofeng Zhu, Xuefeng (Peter) Wang, Xiangqing Sun.}\label{f5}\vspace*{6pt}
\end{figure*}

These days I spend a lot of time writing grants. It's getting harder to get
them. Renewal of my research grants and the training grant is taking more
time than I wish. It's hard to know how much detail the reviewers want.
Sometimes the projects described in the grant proposal get published before
the grant gets funded! And of course, if we don't get funding, we won't be
able to support Ph.D. students to do research.

I also spend a lot of time helping others. For junior faculty, I tell them
not to put my name on their paper as an author because when they come up
for tenure, people may think the paper was my idea and not theirs. If my
name is on a paper, you can be sure I really contributed something. I don't
notice all the authors when I am reading a paper, but I find that people
notice if my name is on a paper. So I have to be sure that every sentence
is accurate. My purpose is to be pedagogical as well as do research, and
this makes me very fussy about proper wording and clarity. I read the
galley proofs personally and I have my secretary read them, too. I learned
that at Aberdeen from Finney. He had a sign up in the tea room which said,
``No paper leaves this department without the Professor's permission.''

I still work on family studies, although that has become less fashionable
than case-control GWAS. My recent work is still a mixture of theoretical
and applied and I still enjoy writing and publishing with students.

\textbf{Gang:} Do you ever plan to retire?

\textbf{Robert:} I don't know when I shall retire---probably when I am no
longer able to get grants to fund my research. I have four children and ten
grandchildren. By the way, three of my children are university professors
in mathematics/health sciences, and the one who isn't decided to be
creative and studied acting. She puts on the high school play every year
(and her older son is majoring in mathematics at college). I look forward
to spending more time with my family when I do \mbox{retire}.

\section{Summing Up}

\textbf{Gang}: You have directed 40 Ph.D. theses and had 45 post-docs. By
now they too have had trainees. What does your ``research pedigree'' look
like?

\textbf{Robert:} My research pedigree has more than 500 progeny The
International Genetic Epidemiology Society had a special tribute for me on
my 70th birthday and someone drew it out. At that time, half of the
field of genetic epidemiology was in my pedigree. There are at least four
generations.

\textbf{Gang:} How would you sum up your career?

\textbf{Robert:} Like that of many other academics, my career path was an
accident. When I talk to others in academia, most of the time they had no
idea what field they would end up in. In my case, I didn't even expect to
go to academia. But once I decided to apply statistics to genetics, I think
I made a happy choice; and I'm glad I decided early on to make all of my
students collaborators. I only hope they learned as much from me as I from
them.

\textbf{Gang:} What is your advice to a young statistical geneticist
starting out today?

\textbf{Robert:} My advice is quite simple. First, make sure you keep
learning as much statistics as you can and second, keep up to date with
computing technology. Statistical genetics may go out of fashion, but there
will always be a need for statisticians who can compute.

\textbf{Zhaohai:} Do you have any closing comments?

\textbf{Robert:} It is really nice to have this conversation appear in
\textit{Statistical Science}. I actually never considered myself to be a
statistician. I was a geneticist among the statisticians and a statistician
among the geneticists!





\end{document}